\documentclass[prl,floatfix,twocolumn,showpacs,superscriptaddress]{revtex4}
\usepackage{graphicx}
\def\bea{\begin{eqnarray}}
\def\eea{\end{eqnarray}}
\def\be{\begin{equation}}
\def\ee{\end{equation}}

\def\S{\mbox{\bf S}}
\def\tJ{$t{-}J$ }
\def\et{{\it et al.}}

\begin{document}

\author{Didier Poilblanc}

\affiliation{
  Laboratoire de Physique Th\'eorique, Universit\'e Paul Sabatier,
  F-31062 Toulouse, France }

\affiliation{
Institute of Theoretical Physics,
Ecole Polytechnique Federale de Lausanne,
BSP 720,
CH-1015 Lausanne,
Switzerland}

\date{\today}
\title{Stability of inhomogeneous superstructures from
           renormalized mean-field theory
of the \tJ Model}

\begin{abstract}
Using the \tJ model (which can also include Coulomb repulsion)
and the ``plain vanilla''
renormalized mean-field theory of Zhang et al. (1988),
stability of inhomogeneous $4a\times 4a$ superstructures as those
observed in cuprates
superconductors around hole doping 1/8 is investigated.
We find a non-uniform $4a\times 4a$ bond order wave
involving simultaneously small ($\sim 10^{-2}t$) inhomogeneous
staggered plaquette currents
as well as a small charge density modulation similar to
pair density wave order.
On the other hand, no supersolid phase involving
a decoupling in the superconducting
particle-particle channel is found.

\end{abstract}
\maketitle

Although the global phase diagram of high-Tc cuprate superconductors
seems, at first sight, to contain only three homogeneous phases,
antiferromagnetic, superconducting  and metallic phases, a closer
inspection reveal a striking complexity with
some form of local electronic ordering around $1/8$ hole doping.
For exemple, recent scanning tunneling microscopy/spectroscopy (STM/STS)
experiments of under-doped
Bi$_2$Sr$_2$CaCu$_2$O$_{8+\delta}$  in the pseudogap state
have shown evidence of real-space modulations of the
low-energy density of states (DOS)~\cite{STM-BSCO} with period
close to four lattice spacings.
Spatial variation of the electronic states
has also been observed in the pseudogap phase of
Ca$_{2-x}$Na$_x$CuO$_2$Cl$_2$ single crystals ($x=0.08\sim0.12$)
by similar STM/STS techniques~\cite{STM2}.
In the same material, signatures of this unusual $4a \times 4a$ checkerboard
charge-ordered state have been found in
Angular Resolved Photoemission Spectroscopy (ARPES) experiments~\cite{ARPES}.
How this local ordering could be explained
on theoretical grounds is a topic of intense ongoing investigation.

In the Resonating Valence Bond (RVB) theory~\cite{RVB}, one of the
early approach to understand strongly correlated HTC
superconductors, a simple variational Ansatz is constructed from a
BCS paired superconducting state $|\Psi_{\rm BCS}\big>$ by a
Gutzwiller~\cite{Gutzwiller} projection of the high energy
configurations with doubly occupied sites. At half-filling, this
procedure generates a highly correlated wavefunction
$P_G|\Psi_{\rm BCS}\big>$ where the spins are paired up in singlet
bonds to form a quantum spin liquid with short range
antiferromagnetic correlations. In fact, due to a SU(2)
electron-hole symmetry, the RVB state can take various forms like
the half-flux state~\cite{half-flux} which can be mapped onto free
electrons on a lattice experiencing half a flux quantum per
plaquette.

The main difficulty to deal with such wavefunctions is to treat
correctly the Gutzwiller projection. Exact Variational Monte Carlo
(VMC) calculations~\cite{VMC} on large clusters show that the
magnetic energy of the variational RVB state at half-filling is
very close to the best exact estimate. It also provides, at finite
doping, a semi-quantitative understanding of the phase diagram of
the cuprate superconductors and of their experimental properties.
Another route to deal with the Gutwiller projection is to use a
``renormalized mean-field (MF) theory''~\cite{Renormalized_MF} in
which the kinetic and superexchange energies are renormalized by
different doping-dependent factors $g_t$ and $g_J$ respectively.
As emphasized recently by Anderson and coworkers~\cite{RVB2}, the
procedure consists in searching
 for an effective MF Hamiltonian that determines a function $|\Psi\big>$ {\it to be projected}
as $P_G|\Psi\big>$. Crucial, now well established, experimental
observations such as the existence of a pseudo-gap and nodal
quasi-particles and the large renormalization of the Drude weight
are remarkably well explained by this early MF theory~\cite{RVB2}.

Away from half-filling, ``commensurate flux phases'' have also
been proposed as another class of competitive projected
wavefunctions~\cite{flux_phases,flux_phases_2}. Such time reversal
symmetry broken states, generalizing Affleck-Marston 1/2-flux
phase away from half-filling, are made out of orbitals solving a
fictitious Hofstadter problem~\cite{hofstadter} of free particles
under a (fictitious) magnetic flux. Within this class of
wavefunctions, the lowest energy is obtained when the flux per
unit cell equals exactly the filling fraction $\nu=\frac{1}{2}(1-x)$
($x$ being the hole
doping)~\cite{flux_phases_2,flux_phases_numerics} suggesting that
doping induces a (scalar) chirality of the spin
background~\cite{chiral_SL}. This scenario also emerges naturally
within a renormalized MF theory~\cite{Renormalized_MF_2} although
the stability condition (self-consistency) leads to a more complex
one-body effective MF Hamiltonian with inhomogeneous flux. For
example, at filling $\nu=p/q$, diagonal ``stripe''
modulations of supercell $qa/\sqrt{2}\times \sqrt{2}a$ ($a$ is the
lattice spacing) are found.
In this Letter, we generalize the MF
approach of Ref.~\cite{Renormalized_MF_2} to allow for non-uniform
densities (and also include off-diagonal pairing). We determine
under which conditions a $4a\times 4a$ superstructure might be
stable for hole doping close to $1/8$.

The weakly doped antiferromagnet
is described here by the renormalized
$t{-}J$ model Hamiltonian,
  \begin{equation}
   H= - tg_t\sum_{\langle ij\rangle\sigma}
      (c^{\dagger}_{i,\sigma}c_{j,\sigma}+h.c.)
    + Jg_J\sum_{\langle ij\rangle}\S_i \cdot \S_j
  \end{equation}
where the local constraints of no doubly occupied sites
are replaced by statistical Gutzwiller weights $g_t=2x/(1+x)$ and $g_J=4/(1+x)^2$.
We shall assume a typical value of $t/J=3$ hereafter.

Decoupling in both particle-hole and (singlet) particle-particle
channels can be considered simultaneously leading to the following
MF hamiltonian,
  \begin{eqnarray}
   H_{\rm MF}= - t \sum_{\langle ij\rangle\sigma} g_{ij}^t
      (c^{\dagger}_{i,\sigma}c_{j,\sigma}+h.c.)-\mu\sum_{i\sigma}n_{i,\sigma}\nonumber \\
   -\frac{3}{4} J \sum_{\langle ij\rangle\sigma}g_{i,j}^J
(\chi_{ji}c^{\dagger}_{i,\sigma}c_{j,\sigma} + h.c. -|\chi_{ij}|^2)\\
   -\frac{3}{4} J \sum_{\langle ij\rangle\sigma}g_{i,j}^J
(\Delta_{ji}c^{\dagger}_{i,\sigma}c^\dagger_{j,-\sigma}
   + h.c. -|\Delta_{ij}|^2),\nonumber
  \end{eqnarray}
where the previous Gutzwiller weights have been expressed
in terms of local fugacities $z_i=2x_i/(1+x_i)$ ($x_i$ is the local hole density),
$g_{i,j}^t=\sqrt{z_i z_j}$ and
$g_{i,j}^J=(2-z_i)(2-z_j)$, to allow for small non-uniform charge modulations~\cite{Anderson2004}.
The Bogolubov-de Gennes self-consistency conditions are implemented as
$\chi_{ji}=\langle c^\dagger_{j,\sigma}c_{i,\sigma}\rangle$ and
$\Delta_{ji}=\langle c_{j,-\sigma}c_{i,\sigma}\rangle
=\langle c_{i,-\sigma}c_{j,\sigma}\rangle$.

Assuming an homogeneous system, the RVB theory of
Refs.~\cite{RVB,RVB2} is recovered leading to a $d_{x^2-y^2}$ RVB
superconducting (for $x>0$) ground state (GS), $\Delta_{i\
i+x}=\Delta$, $\Delta_{i\ i+y}=-\Delta$ and $\chi_{ij}=\chi$. When
doping $x\rightarrow 0$, $\Delta=\chi\simeq 0.169$, giving a
magnetic energy per bond of $\sim 0.339 J$ and a kinetic energy
per hole of $\sim -2.71 t$ (within MF). $J\Delta$ is physically
the pseudo-gap while the {\it superconducting} temperature scales
like $g_tJ\Delta\propto x$.

Our aim here is to investigate inhomogeneous solutions of these
equations which  might {\it a priori} break translation symmetry.
However, searching for the global energy minimum in a completely
unconstrained variational space is a formidable task owing to the
non-linear nature of the MF equations. Therefore, guided by
experimental observations, for an average hole density of $x\simeq
1/8$, we shall restrict ourselves to a $4a\times 4a$
superstructure (which could, as well, accommodate a  smaller
spatial periodicity) with a natural $C_4$ rotation symmetry, hence
naturally restricting the number of non-equivalent bonds and sites
to 6 and 3, respectively. Within these restrictions, MF equations
are solved on finite clusters (with periodic boundaries)  of sizes
ranging from $16\times 16$ to $48\times 48$.

\begin{figure}
  \centerline{\includegraphics*[angle=0,width=0.75\linewidth]{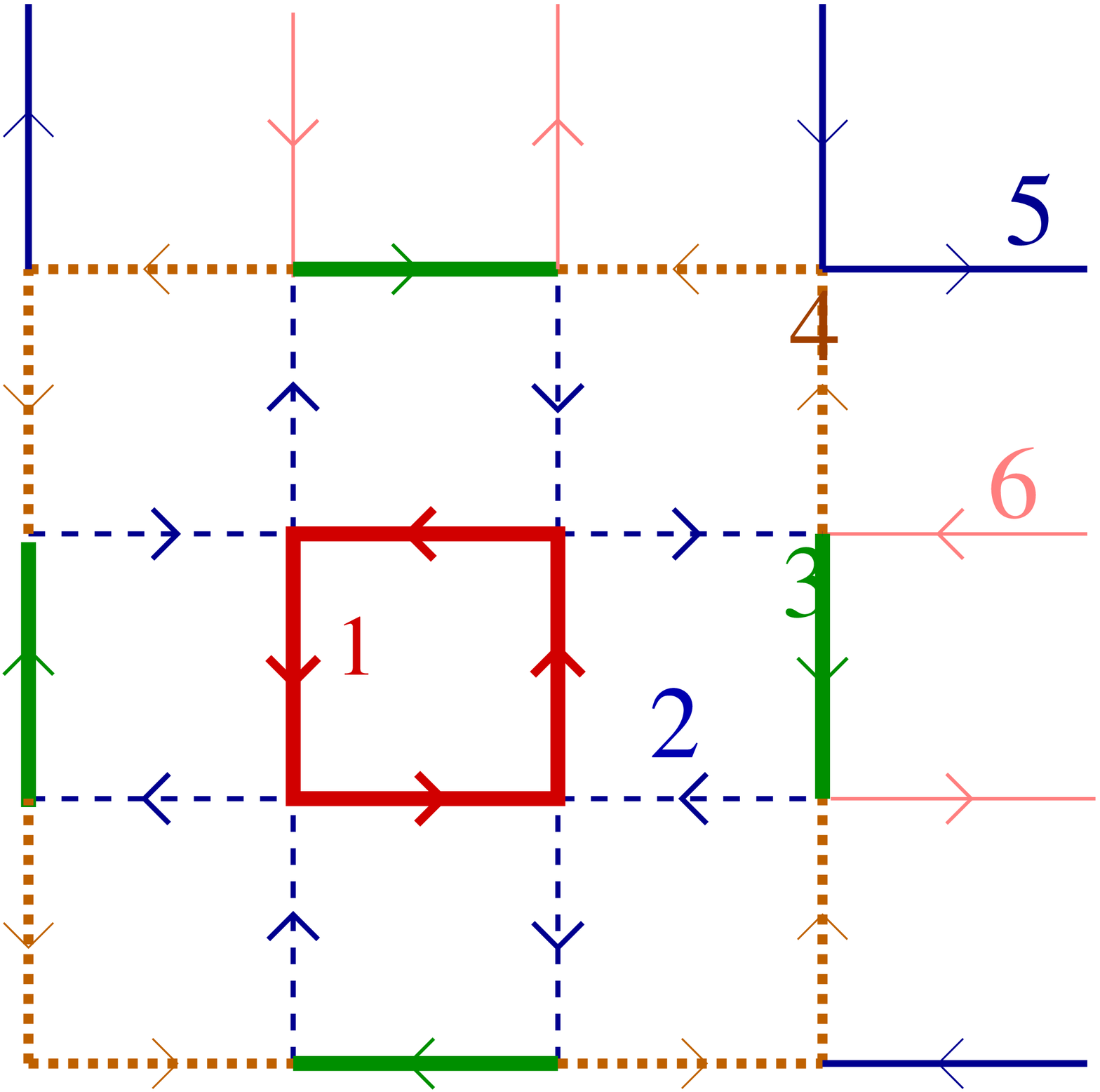}}
  \caption{\label{fig:pattern4x4}
(Color on-line). Schematic pattern of the bond $\chi_{ij}$
distribution. Non-equivalent bonds (labelled from 1 to 6) are
indicated by different types of lines whose widths qualitatively
reflect the magnitudes of $\chi_{ij}$. Arrows indicate the
directions of the small charge currents. 
Numerical values for $t/J=3$ are given in
Table~\protect\ref{table:flux}.}
\end{figure}

We first consider non-superconducting solutions i.e.
$\Delta_{ij}=0$ on all bonds. Starting with an initial set of
$\chi_{ij}$ which simultaneously (i) corresponds to a uniform
(fictitious) flux (per plaquette) equal to the filling fraction $\nu=7/16$
and (ii) fulfills the above symmetry requirements within the
supercell, we then solve the self-consistent equations
iteratively~\cite{note2}. A typical MF solution is shown
schematically in Fig.~\ref{fig:pattern4x4}. As shown in
Table~\ref{table:flux}, the bond parameters $\chi_{ij}$, here
obtained e.g. on a $48\times 48$ cluster~\cite{note3}, show strong
spatial modulations leading to a bond order wave (BOW) in both
spin-spin correlations $\frac{3}{2}g_{ij}^J|\chi_{ij}|^2$ and
charge hoppings $2g_{ij}^t{\rm Real}\{\chi_{ij}\}$. At first
sight, such a state could be seen as an extension of the
bond-centered stripe~\cite{Vojta} or $4\times 4$
plaquette~\cite{Vojta2} solutions found in some $SU(2N)$/$Sp(2N)$
mean field theories. However, in our case, $\chi_{i,j}$ exhibit
imaginary parts (as in staggered~\cite{staggered_flux} and
commensurate flux states) producing a time-reversal symmetry
broken state and small plaquette charge currents $2g_{ij}^t{\rm
Im}\{\chi_{ij}\}$ (see Table~\ref{table:flux}). The 
current pattern is similar to the alternating distribution 
of the staggered flux state 
although the amplitude of the current loops is not uniform.
As shown on the first
line of Table~\ref{table:charge}, such a state also exhibits a
small charge density wave (CDW) component.

\begin{table}[htb]
\caption{\label{table:flux}
Numerical values of the bond amplitudes $|\chi_{ij}|$, the bond spin-spin
correlations
and the bond average hoppings and currents
(in units of $t$)
in the normal phase MF solution obtained on a $48\times 48$ cluster
for $t/J=3$ and $2022$ fermions on $48^2$ sites ($x\simeq 1/8$). Note
the current conservation
at the lattice sites, in particular ``2+6''=''3+4''.}
\begin{center}
\begin{tabular}{|l|l|l|l|l|l|l|}
\hline
Bond $\#$ & 1& 2& 3&   4&   5& 6 \\
\hline
$|\chi_{ij}|$ & 0.2579 & 0.1871 & 0.2595 &  0.2259 & 0.1887 & 0.1639\\
$\big<{\bf S}_i\cdot{\bf S}_j\big>$
& -0.3456 & -0.1720 & -0.3131 & -0.2359 &  -0.1637 & -0.1249\\
hoppings & 0.0554 & 0.0571& 0.1188 & 0.1063 & 0.0915  &0.0725\\
currents & 0.0453& 0.0373 & 0.0370 & 0.0305 & 0.0228  &0.0302\\
\hline
\end{tabular}
\end{center}
\end{table}

We now turn to superconducting solutions allowing pair order
parameter $\Delta_{ij}\ne 0$. We repeat the previous procedure
adding to $\chi_{ij}$ arbitrary initial values for $\Delta_{ij}$,
only requesting $|\Delta_{ij}|$ to have s-wave symmetry w.r.t the
{\it center of the central plaquette} (defined by bonds $\#$1).
Within this restriction, we have looked for 3 candidates
compatible with a $4a\times 4a$ cell, exhibiting (i)
$d_{x^2-y^2}+is$, (ii) $d_{x^2-y^2}+id_{xy}$ or (iii)
$d_{x^2-y^2}+id_{x^2-y^2}$ symmetries. Note that we allow again
inhomogeneous solutions so that the minimization of the MF energy
w.r.t.the set of $\Delta_{ij}$ is performed indeed on 6
independent complex parameters. However, for a doping around $1/8$
and all cluster studied, no modulated solution was found. On the
contrary, the iteration procedure always converges towards the
standard $d_{x^2-y^2}$ RVB superconductor, a state particularly
rigid w.r.t. the establishment of spontaneous modulations.

\begin{figure}
  \centerline{\includegraphics*[angle=0,width=0.75\linewidth]{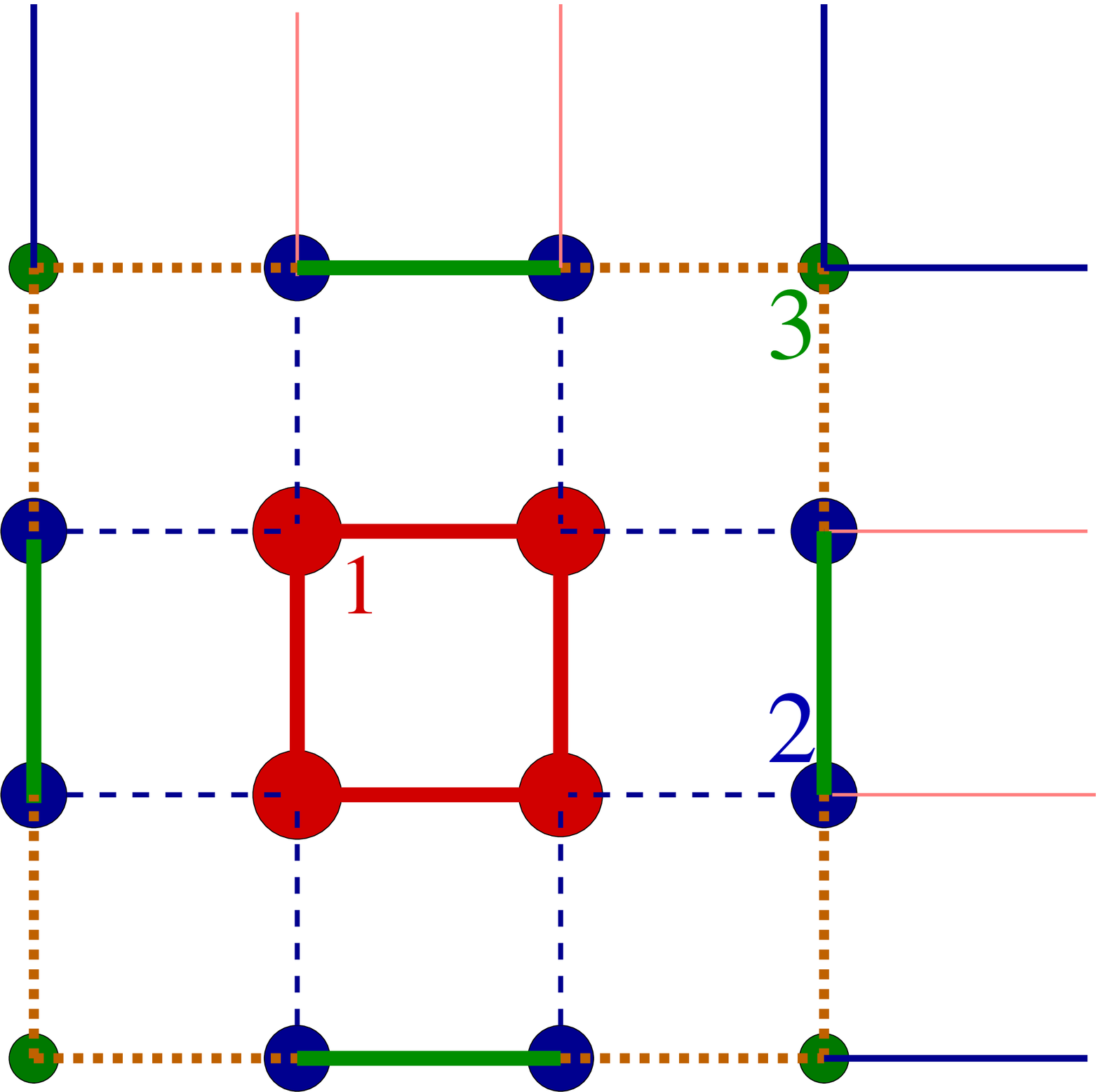}}
  \caption{\label{fig:charge4x4}
(Color on-line). Schematic pattern of the charge and bond
modulation. Non-equivalent sites (labeled from 1 to 3) and bonds
are indicated by different types of circles and lines (and colors,
online). Numerical values for $t/J=3$ are given in
Table~\protect\ref{table:charge}.}
\end{figure}

Next, we investigate the role of a Coulomb repulsion,
  \begin{equation}
   H_V= \frac{1}{2}\sum_{i,j}V_{i,j}(n_i-n)(n_j-n)\, ,
  \end{equation}
that we treat within a simple decoupling scheme, i.e. replacing
the density-density interaction by $\big<n_i\big>n_j+
\big<n_j\big>n_i+{\rm Cst}$. We assume here both long range
$V_{i,j}=V/r_{i,j}$ and screened $V_{i,j}=V^\prime
\exp{(-r_{ij}/l)}/r_{i,j}$ potentials (with a screening length set
to $l=4a$). To minimize finite size effects, we also use the
"periodic  distance" $r_{ij}$ on the torus. As shown in
Table~\ref{table:charge}, we find that moderate values of $V$ (or
$V^\prime$) $\sim t$ affects only slightly the form of the CDW.
Moreover, the pattern for both spin-spin correlations and
effective hoppings remains qualitatively similar to $V=0$ (see
Table~\ref{table:flux2}). Note that the charge current magnitude
tends to increase (decrease) with $V$ on the central, (outer)
plaquette, i.e. on bonds $\# 1$ ($\# 5$). As expected, results do
not depend drastically on the long-distance part of $V_{i,j}$.

\begin{table}[htb]
\caption{\label{table:charge} Numerical values of charge densities
in the normal phase obtained on a $48\times 48$ cluster for
$t/J=3$ and doping $x\simeq 1/8$ and different Coulomb repulsions.
Average density, magnetic bond energy, kinetic energy per hole and
energy per site are provided in the 4 last columns. Last line:
same data obtained for the d-wave RVB superconductor on a
$32\times 32$ cluster for $\mu=-0.4$ (leading to $\chi\simeq
0.192$ and $\Delta\simeq 0.120$).}
\begin{center}
\begin{tabular}{|l|l|l|l|l|l|l|l|}
\hline
Site $\#$ & 1& 2& 3 & $n$&  $\big<{\bf S}_i\cdot{\bf S}_j\big>$ & $E_K/t$  & $e_0/J$  \\
\hline
$V=0$ & 0.925& 0.864 & 0.857 & 0.8776& -0.220  & -2.660  &-1.438 \\
$V/t=0.5$ & 0.894 & 0.913 & 0.783 &0.8759& -0.227  & -2.588  &-1.425 \\
$V/t=0.8$ & 0.876& 0.940 & 0.760&0.8793& -0.232  & -2.396 &-1.367 \\
$V^\prime/t=0.8$ & 0.850& 0.932 & 0.786 &0.875& -0.222  & -2.559 &-1.406 \\
${\rm RVB}$ & 0.876 & 0.876 & 0.876 & 0.8762& -0.243  & -2.730  &-1.500 \\
\hline
\end{tabular}
\end{center}
\end{table}

Let us now turn to the issue of the relative stability of the BOW/CDW state
w.r.t. the RVB superconductor. For $V=0$ the latter has a slightly
lower energy. However, a moderate repulsion can easily stabilize the
BOW/CDW state whose energy is only weakly affected by $V$ (or $V^\prime$).
On the contrary, the pairing amplitude of the
RVB superconductor is expected to be quite sensitive to the Coulomb potential.
Indeed, a decoupling in
the particle-particle channel of e.g. the nearest neighbor
repulsion $V_1 n_{i\uparrow}n_{j\downarrow}$ term
generates a positive contribution $\propto V_1 |\Delta_{ij}|^2$ which tends
to strongly reduce the pairing amplitude. Therefore, the BOW/CDW state
is expected to become stable when $x\sim 1/8$.

\begin{table}[htb]
\caption{\label{table:flux2} Same as
Table~\protect\ref{table:flux} but for a screened Coulomb
potential $V'/t=0.8$ ($l=4a$). $2016$ fermions on $48^2$ sites
($x=1/8$).}
\begin{center}
\begin{tabular}{|l|l|l|l|l|l|l|}
\hline
Bond $\#$  & 1& 2& 3&   4&   5& 6 \\
\hline
$|\chi_{ij}|$ & 0.2297 & 0.2015 & 0.2741 &  0.1917 & 0.2166 & 0.2020\\
$\big<{\bf S}_i\cdot{\bf S}_j\big>$
& -0.2395 & -0.1984 & -0.3952 & -0.1700 &  -0.1909 & -0.2146\\
hoppings & 0.1105 & 0.0637& 0.0547 & 0.0773 & 0.1524  &0.0403\\
currents & 0.0459& 0.0366 & 0.0436 & 0.0251 & 0.0109  &0.0321\\
\hline
\end{tabular}
\end{center}
\end{table}

Lastly, we would like to discuss similarities and differences of
our findings with previous theoretical suggestions to explain the
$4a\times 4a$ patterns experimentally observed e.g. in BiSCO.
First, we should stress that the state found in this work is quite
different from an ordinary CDW or Wigner solid which, for a doping
around $1/8$, is expected to have a $\sqrt{8}\times\sqrt{8}$
supercell. As a matter of fact, the BOW {\it is not} primarily
stabilized by Coulomb repulsion. Secondly, we notice that it bears
some similarities (in addition to the common $4a\times 4a$
periodicity) with the insulating Pair density Wave (PDW)
state~\cite{pdw} which also does not exhibit $U(1)$ phase
coherence. For exemple, it shows a higher hole density on a given
plaquette within the supercell (which, incidently, carries the
minimum value of the current). Like the PWD state, the BOW differs
from the stripe state in terms of the rotational symmetry. In
addition, in contrast to PDW and stripe states, it breaks
time-reversal symmetry. However, an experimental observation of
its manifestation i.e. the plaquette orbital currents might be
quite tedious due to their small magnitudes. We note that the
current pattern is, at first glance, staggered (in fact,
incommensurate with a wave vector close to $(\pi,\pi)$), a finding
consistent with exact diagonalisations of doped Hubbard and \tJ
clusters showing enhancement of {\it staggered} chiral
correlations under doping~\cite{ED}. Finally, we observe that
recent VMC calculations of staggered flux
states~\cite{staggered_flux} suggest an instability towards phase
separation. A modulated structure like the one found here is in
fact an alternative scenario.

In conclusion, using renormalized meanfield theory of the \tJ model,
$4a\times 4a$ BOW states are found for doping around $1/8$.
We suggest that such states might be relevant to explain the
spatial modulations of the DOS in STM experiments of underdoped
cuprates superconductors.
They carry small (almost staggered) plaquette currents and
are stable for moderate Coulomb repulsion.

I thank IDRIS (Orsay, France) for CPU-time as well as the
Institute for Theoretical Physics (EPFL, Switzerland) for
hospitality. I also acknowledge useful conversations with
S.~Capponi, B.~Kumar, F.~Mila and F.C.~Zhang.


\begin{thebibliography}{}

\bibitem{STM-BSCO} M.~Vershinin, S.~Misra, S.~Ono. Y.~Abe, Y.~Ando
and A.~Yazdani, Science {\bf 303}, 1995 (2004).

\bibitem{STM2} T. Hanaguri \et, Nature {\bf 430}, 1001 (2004).

\bibitem{ARPES} K.M.~Shen \et, cond-mat/0407002 and Invited talk at
the 2005 APS March meeting.

\bibitem{RVB} P.W.~Anderson, Science {\bf 237}, 1196 (1987).

\bibitem{Gutzwiller} M.C.~Gutzwiller, Phys.~Rev.~Lett. {\bf 10}, 159 (1963);
D.~Vollhardt, Rev.~Mod.~Phys.
{\bf 56}, 99 (1984).

\bibitem{half-flux} I.~Affleck and J.B.~Marston, Phys.~Rev.~B. {\bf 37}, R3774 (1988);
J.B.~Marston and I.~Affleck, {\it ibid.} {\bf 39}, 11538 (1989);
G.~Kotliar, {\it ibid.} {\bf 37}, 3664 (1988).

\bibitem{VMC} C.~Gros,  Phys.~Rev.~B {\bf 38}, R931 (1988);
For recent estimations see e.g.
A.~Paramekanti, M.~Randeria and N.~Trivedi,
Phys.~Rev.~Lett. {\bf 87}, 217002 (2001).

\bibitem{Renormalized_MF} F.C.~Zhang, C.~Gros, T.M.~Rice and H.~Shiba,
Supercond.~Sci.~Technol.~{\bf 1}, 36 (1988).

\bibitem{RVB2} P.W.~Anderson, P.A.~Lee, M.~Randeria, T.M.~Rice, N.~Trivedi
and F.C.~Zhang, J Phys. Condens. Matter {\bf 16}, R755-R769 (2004).

\bibitem{flux_phases} P.W.~Anderson, B.S.~Shastry and D.~Hristopulos,
Phys.~Rev.~B. {\bf 40}, 8939 (1989);
D.~Poilblanc, {\it ibid.} {\bf 40}, R7376 (1989).

\bibitem{flux_phases_2}  P.~Lederer, D.~Poilblanc and T.M.~Rice,
Phys.~Rev.~Lett. {\bf 63}, 1519 (1989);
F.~Nori, E.~Abrahams and
G.T.~Zimanyi, Phys.~Rev.~B. {\bf 41}, R7277 (1990).

\bibitem{hofstadter} D.R.~Hofstadter,  Phys.~Rev.~B. {\bf 14}, 2239 (1976).

\bibitem{flux_phases_numerics} For numerical computations of {\it projected}
flux phases see e.g. D.~Poilblanc, Phys.~Rev.~B. {\bf 39}, 140 (1990);
D.~Poilblanc, Y.~Hasegawa, and T.M.~Rice,
{\it ibid.} {\bf 41}, 1949 (1990).

\bibitem{chiral_SL} Frustration is essential to stabilize chiral spin liquids.
See e.g. X.G.~Wen, F.~Wilczek and A.~Zee,  Phys.~Rev.~B. {\bf 39}, 11413 (1990).

\bibitem{Renormalized_MF_2} D.~Poilblanc, Phys.~Rev.~B. {\bf 41}, R4827 (1990).

\bibitem{Anderson2004}
P.W.~Anderson, cond-mat/0406038; B.A.~Bernevig \et,
cond-mat/0312573.

\bibitem{note2} $n$ is allowed to slightly change
during the procedure to avoid degeneracies at Fermi level.

\bibitem{note3} Comparisons to smaller clusters show that finite size effects
remain quite small.

\bibitem{Vojta} M.~Vojta, Y.~Zhang and S.~Sachdev,
 Phys.~Rev.~B. {\bf 62}, 6721 (2000) and references therein.

\bibitem{Vojta2} M.~Vojta, Phys.~Rev.~B. {\bf 66}, 104505 (2002);
Note that here the large-N Sp(2N) scheme implies
a superconducting state.

\bibitem{staggered_flux}
D.A.~Ivanov, Phys.~Rev.~B. {\bf 70}, 104503 (2004) and references therein;
see also D.~Poilblanc and Y.~Hasegawa, Phys.~Rev.~B {\bf 41}, 6989 (1990).


\bibitem{pdw}
H.-D. Chen, O.~Vafek, A.~Yazdani, and S.-C.~Zhang,
Phys.~Rev.~Lett. {\bf 93}, 187002 (2004);
H.-D.~Chen, S.~Capponi, F.~Alet and S.-C.~Zhang,
Phys.~Rev.B {\bf 70}, 024516 (2004).

\bibitem{ED} D.~Poilblanc, E.~Dagotto and J,~Riera, 
Phys.~Rev.~B {\bf 43}, 7899 (1991); E.~Dagotto et al.,
{\it ibid} {\bf 45}, 10741 (1992).

\end{thebibliography}
\end{document}